\documentclass{article}
\usepackage{amsmath,epsfig}
\usepackage[preprint]{spconfa4}
\usepackage{multirow, subfigure, amssymb, bm}

\let\OLDthebibliography\thebibliography
\renewcommand\thebibliography[1]{
  \OLDthebibliography{#1}
  \setlength{\parskip}{0pt}
  \setlength{\itemsep}{0pt plus 0.3ex}
}

\begin{document}\sloppy

% Example definitions.
% --------------------
\def\x{{\mathbf x}}
\def\L{{\cal L}}

% Title.
% ------
\title{TASK-RELATED SELF-SUPERVISED LEARNING FOR REMOTE SENSING IMAGE CHANGE DETECTION}
%
% Address.
% ---------------
\name{Zhinan Cai, Zhiyu Jiang$^*$, Yuan Yuan\thanks{2021 IEEE. Personal use of this material is permitted. Permission from IEEE must be obtained for all other uses, in any current or future media, including reprinting/republishing this material for advertising or promotional purposes, creating new collective works, for resale or
		redistribution to servers or lists, or reuse of any copyrighted component of this work in other works. $^*$Corresponding author: Zhiyu Jiang (e-mail: zhiyu.jiang.chn@gmail.com).}}
%\name{XXX, XXX, XXX$^*$\thanks{}}
%\address{School of Computer Science and Center for OPTical IMagery Analysis and Learning (OPTIMAL),	\\ Northwestern Polytechnical University, Xi’an, Shaanxi, P.R. China, 710072.}
\address{School of Computer Science and School of Artificial Intelligence, Optics and Electronics (iOPEN), \\Northwestern Polytechnical University, Xi'an 710072, P.R. China}

\address{School of Computer Science and School of Artificial Intelligence, Optics and Electronics (iOPEN),\\
Northwestern Polytechnical University, Xi'an 710072, P.R. China\\
z.cai@mail.nwpu.edu.cn; y.yuan1.ieee@gmail.com; jiangzhiyu@nwpu.edu.cn}

%\address{yuejiao@mail.nwpu.edu.cn, y.yuan1.ieee@gmail.com, jiangzhiyu@nwpu.edu.cn}

\maketitle

\begin{abstract}
	Change detection for remote sensing images is widely applied for urban change detection, disaster assessment and other fields. However, most of the existing CNN-based change detection methods still suffer from the problem of inadequate pseudo-changes suppression and insufficient feature representation. In this work, an unsupervised change detection method based on Task-related Self-supervised Learning Change Detection network with smooth mechanism(TSLCD) is proposed to eliminate it. The main contributions include:
	(1) the task-related self-supervised learning module is introduced to extract spatial features more effectively.
	(2) a hard-sample-mining loss function is applied to pay more attention to the hard-to-classify samples.
	(3) a smooth mechanism is utilized to remove some of pseudo-changes and noise.
	Experiments on four remote sensing change detection datasets reveal that the proposed TSLCD method achieves the state-of-the-art for change detection task.
\end{abstract}
\section{Introduction}
\label{sec:intro}

Change detection plays an important role in human interpretation of remote sensing images. On the one hand, change detection task can promptly and accurately detect the areas that have changed in the land. On the other hand, it provides a basis for land  management and predicting change trends. Its application fields and research scopes are also very wide, including areas such as land cover change detection \cite{9053840}, forest cover detection \cite{8683128}, wetland change detection \cite{ruan2007change}, urban area expansion \cite{8683526}, disaster assessment \cite{gong2013earthquake}, and coastline changes \cite{valderrama2019dynamics}.

A lot of techniques have been developed to detect changes from multitemporal remote sense images over the last decades. Most of these methods can be divided into two categories: supervised and unsupervised \cite{zhang2018coarse}. Among the classical supervised methods, Mou \emph{et al.} \cite{mou2018learning} proposed ReCNN, which is trained to extract  a joint spectral–spatial–temporal feature representation. Zhao \emph{et al.} \cite{zhao2019incorporating} proposed MeGAN to explore seasonal invariant features for pseudo-changes suppressing. Zhang \emph{et al.} \cite{9052762} pretrained the network with remote sensing image scene classification dataset, and then utilized transfer learning and a shared-weight-network to generate a multiscale feature difference map for change detection. Those methods require prior knowledge and tremendous number of manually labeled samples, otherwise the model will be less generalized and robust \cite{wei2020detection, jiang2018contour, wang2020caption}. 
%The FDCNN method 
%can obtain a better feature representation as pretraining on large scale data has 
%better representational capability than random initialization.
%However, the features obtained by FDCNN method are not consistent with 
%the features required by the change detection task.

\begin{figure*}[htb]
	\begin{minipage}[b]{1.0\linewidth}
		\centering
		%		\centerline{\includegraphics[width=18cm]{image5}}
		\centerline{\includegraphics[width=\textwidth]{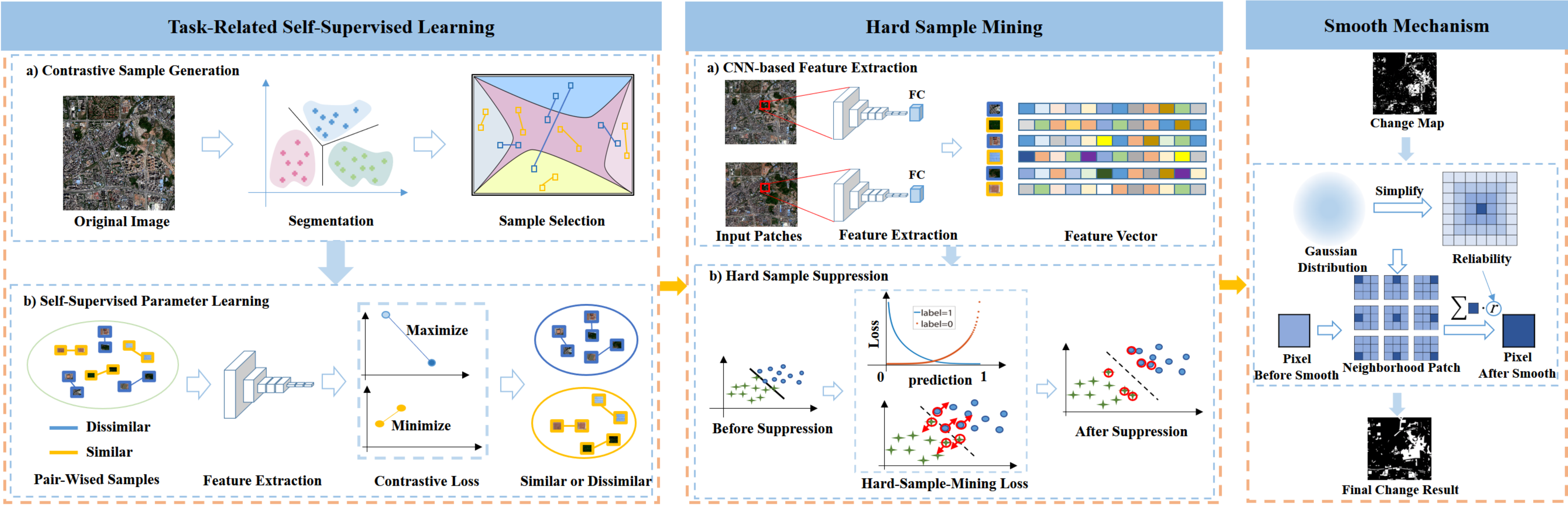}}
		%  \vspace{2.0cm}
		%\centerline{() }\medskip
	\end{minipage}
	\vspace{-0.8cm}
	\caption{The structure of proposed Task-related Self-supervised Learning Change Detection network (TSLCD), which is composed of task-related self-supervised module, hard-sample-mining loss function and smooth mechanism.}
	\label{fig:res1}
	\vspace{-0.5cm}
\end{figure*}

%Celik \cite{celik2009unsupervised} utilized Principal Component Analysis method to detection changes in multitemporal satellite images, which can reduce data redundancy, leaving only information that can show differences.

To solve this problem, many unsupervised change detection methods have been proposed in the past decades. Saha \emph{et al.} \cite{saha2019unsupervised} proposed DCVA to obtain the difference image by generating the spectral change vector,  and to obtain the final change map by threshold-based or clustering-based approaches. Han \emph{et al.} \cite{4063292} proposed K-T Transformation, which is based on the information distribution structure in the multi-dimensional spectral space to perform an empirical linear orthogonal transformation on the image. Do \emph{et al.} \cite{du2019unsupervised} used two deep networks to extract features and utilized Slow Feature Analysis to extract the most invariant components of unchanged pixels. However, most of the above approaches have their limitations, and their change detection results generally have a lot of pseudo-changes, which accounts for high false alarm rate. Generally, pseudo-changes refer to shadow changes and vegetation color changes caused by weather and seasonal variation.

In this work, a Task-related Self-supervised Learning-based Change Detection network with smooth mechanism(TSLCD) is proposed, which can obtain a task-related feature representation and remove more pseudo-changes without any manually labeled samples. The contributions of this work are summarized as follows:

%	(1) A self-supervised learning module related to change detection tasks is proposed to obtain a better pretrain network without any labeled samples.
%By introducing this unit, a better feature representation highly consistent with ones required by the change detection task can be obtained.
(1) In order to obtain task-related features without manually labeled samples, a self-supervised learning module is proposed to generate a large number of labeled similar and dissimilar pair-wised samples, which corresponds to the unchanged and changed pair-wised samples in change detection task.

%	(2) A hard-samples-mining loss function is proposed to change the weight of those hard-to-classify samples, thereby increasing their contribution to the loss function.
%This makes the model pay more attention to the hard-to-classify samples, so as to get better classification results.
(2) For mining those hard-to-classify samples, a hard-sample-mining loss function is proposed to change the weight of those hard-to-classify samples, which allows the network to pay more attention to hard-to-classify samples.

(3) To remove some of noise and pseudo-changes, which refer to shadow changes and vegetation color changes caused by weather and seasonal variation, smooth mechanism is proposed to obtain the reliability of the change detection results.

\vspace{-0.25cm}
\section{PROPOSED FRAMEWORK}
\label{sec:framework}

The overall architecture of the proposed framework is illustrated in Fig. 1. Logically, the approach consists of three main steps. First, a task-related self-supervised module is trained to extract deep features consistent with change detection task in a large number of negative and positive pair-wised samples. Second, based on the proposed hard-sample-mining loss function, TSLCD is trained by sharing parameters model with the trained self-supervised network. Third, based on the change map generated by TSLCD network, the smooth mechanism can be utilized to remove some of noise and pseudo-changes, which refer to shadow changes and vegetation color changes caused by weather and seasonal variation.

%\begin{figure}[htb]
%	\begin{minipage}[b]{1.0\linewidth}
%		\centering
%		\centerline{\includegraphics[width=8.4cm]{image6}}
%		%  \vspace{2.0cm}
%		%\centerline{(a) Result 1}\medskip
%	\end{minipage}
%	\caption{The principle of smooth mechanism.}
%	\label{fig:res2}
%\end{figure}
\vspace{-0.2cm}
\subsection{Task-Related Self-Supervised Module}	
\label{ssec:details}

For the sake of understanding, this work uses ${{{\bm{X}}}_{t0}},{{{\bm{X}}}_{t1}} \in {\mathbb{R}^{H \times W \times 3}}$ to represent the first and second period remote sensing images respectively. The ground truth and the change result map are represented by ${\bm{Y}_{gt}} \in {\{ 0,1\} ^{H \times W}}$ and $\bm{Y}_{rm} \in {\{ 0,1\} ^{H \times W}}$.

The whole task-related self-supervised module consists of two stages, pair-wised samples selection and model training. In order to obtain better feature representation, it is usually necessary to use supervised learning, which requires sufficient manually labeled samples. However, manually labeling data is a time-consuming and expensive process. Therefore, this work designs a task where it can generate a large number of labels from existing images and use these labels to learn the representation of the image. To make the feature representation obtained by the  model consistent with the feature representation obtained by the change detection task, it generates a large number of similar and dissimilar pair-wised samples. For the change detection task, the similarity value of changed pair-wised samples is smaller, and the similarity value of the unchanged pair-wised samples is bigger. In order to generate  similar and dissimilar pair-wised samples, this work uses an unsupervised image segmentation algorithm \cite{8296898} to segment the original image, which can be described as:
\begin{equation}
	{\bm{X}_s} = {f_{seg}} ( {{\bm{X}}_{t0}},{{\bm{X}}_{t1}},{\rm{k}} ),
\end{equation}
where ${\bm{X}_s}$ is the segmented image and ${k}$ is the number of classes. We believe that the between-class samples are not similar, and the within-class samples are similar. Therefore, a reliable sampling strategy is utilized to select $m_s$ pairs of similar pair-wised samples and $m_d$ pairs of dissimilar pair-wised samples from the segmentation map, which can be described as:
\begin{equation}
	%	\frac{{{\rm{S(}}{x_\xi } \in {N_{ij}} \cap {{\rm{K}}_\xi }{\rm{ = }}{{\rm{K}}_{ij}}{\rm{)}}}}{{n \times n}} \ge P,
	%	\frac{{{\rm{S(}}{N_{ij}} \in \{ {{\bm{X}}_{t0}},{{\bm{X}}_{t1}}\}  \cap {{\rm{K}}_{ij}}{\rm{ = }}{{\rm{K}}_\delta }{\rm{)}}}}{{n \times n}} \ge p
	\frac{{|\{ {N_{ij}}|{N_{ij}} \in \mathbb{N},{C_{{N_{ij}}}} = {C_K}\} |}}{{n \times n}} \ge p
\end{equation}
%where the sample ${x_\xi}$ is in the neighborhood of ${N_{ij}}$ with a size of ${n \times n}$, 
%and $S(\cdot)$ means the number of other samples in the neighborhood ${N_{ij}}$ that have the same class with
%${x_\xi}$.
where $\mathbb{N}$ is a patch with size of $n \times n$ cropped from ${\bm{X}_s}$. $K$ is the central pixel of the patch $\mathbb{N}$. ${N_{ij}}$ is the $i$th row and the $j$th column pixel of the patch $\mathbb{N}$. $C_{K}$ is the class of the central pixel $K$. ${C_{{N_{ij}}}}$ is the class of the pixel ${N_{ij}}$. $|\cdot|$ represents the total number of elements in set $\{\cdot\}$, which indicates the number of other samples in the patch $\mathbb{N}$ that have the same class with central pixel. Eq. 2 indicates that when $|\cdot|$ divided by ${n \times n}$ is greater than $p$, the reliable patch $\mathbb{N}$ is selected with a size of ${n \times n}$.

To ensure that the model can effectively exploit both positive and negative pair-wised samples, the size of $m_s$ and $m_d$ are kept at the same order of magnitude. The backbone network is composed of four convolutional layers, where the size of the convolution kernel is 3 × 3. The convolutional layers combines the Rectified Linear Unit (ReLU). we utilized euclidean distance to measure the similarity of samples. The loss function is described as follows:
\begin{equation}
	L_{self} = \frac{1}{{2N}}\sum\limits_{n = 1}^N {y{d^2} + (1 - y)\max {{(m - d,0)}^2}},
\end{equation}
where $d$ is the euclidean distance between the features of pair-wised samples. $m$ is a margin for separating similar and dissimilar pair-wised samples. As stated in Eq. 3, to train the task-related self-supervised module, the contrastive loss is adopted as the loss function.
\vspace{-0.2cm}
\subsection{Hard-Sample-Mining Loss Function}	
\label{ssec:loss}

In the task-related self-supervised learning stage, the feature representation obtained by the self-supervised module is more suitable for change detection task. Due to the high consistency between the designed task and change detection task, the module can make the training stage converge faster and improve the change detection results. Therefore, the backbone network in the change detection stage adopts two weight-sharing pre-trained self-supervised network to extract the features of the input patches, and then cascades the two features into the two fully connected layers to obtain a binary change map.

A cross-entropy loss is adopted as the loss function to train the model. Because the contribution of some easy-to-classify samples and hard-to-classify samples to loss is close during the training process, the model cannot exploit hard-to-classify samples effectively. Therefore, in order to pay more attention to hard-to-classify samples, this work considers mining the information of hard-to-classify samples. As shown in Eq. 4, the work introduce the gamma factor to increase the weight of hard-to-classify samples. By changing the size of gamma, the contribution of hard-to-classify samples to loss can be increased, so that the network pays more attention to the hard-to-classify samples and learns more helpful knowledge for the task.
%\begin{equation}
%	L_{mining} = \left\{ \begin{array}{l}
%	- {\sigma ^\gamma }(0.5 - \mathop y\limits^ \wedge  )\log (\mathop y\limits^ \wedge  ),y = 1\\
%	- {\sigma ^\gamma }(\mathop y\limits^ \wedge   - 0.5)\log (1 - \mathop y\limits^ \wedge  ),y = 0,
%	\end{array} \right.
%	L_{mining} = - y [{\sigma ^\gamma }(0.5 - \mathop y\limits^ \wedge  )]\log (\mathop y\limits^ \wedge  ) \\ \\
%	-(1-y)[{\sigma ^\gamma }(\mathop y\limits^ \wedge   - 0.5)]\log (1 - \mathop y\limits^ \wedge  )\\
\begin{align}
	L_{mining} =  &- y[{\sigma ^\gamma }(0.5 - \mathop y\limits^ \wedge  )]\log (\mathop y\limits^ \wedge  ) \notag \\&- (1 - y)[{\sigma ^\gamma }(\mathop y\limits^ \wedge   - 0.5)]\log (1 - \mathop y\limits^ \wedge  ),
\end{align}
%\end{equation}
where ${\sigma}$ is sigmoid function, ${y \in {\bm{Y}_{gt}}}$ is the label of patch and ${\mathop y\limits^ \wedge   \in {\bm{Y}_{rm}}}$ is the predict of patch. ${\gamma}$ is gamma factor. By increasing the value of gamma factor, the contribution of hard-classify-sample to loss can be increased.
\vspace{-0.2cm}
\subsection{Smooth Mechanism}	
\label{ssec:SM}

The change map obtained above still contains a lot of pseudo-changes and noise, which refer to shadow changes and vegetation color changes caused by weather and seasonal variation. To eliminate these, a smooth mechanism has been proposed. In order to facilitate model learning, the label of a patch's central pixel is used to represent the label of the entire patch. Therefore, for the entire patch, the reliability of each pixel label is negatively related to the distance between the pixel and the central pixel, which is modeled by two-dimensional Gaussian distribution.
%To simplify the problem, the label's reliability of the $n \times n$ patch is divided into $(n+1)/2$ categories according to the distance.

As shown in Fig. 1, the label's reliability of each pixel in a patch decreases as the distance from the center pixel increases. The label's reliability of each pixel is obtained by accumulating the reliability of all patches who contains the pixel, which is described as follows:
\begin{equation}
	%\left\{ \begin{array}{l}
	%{x_{j}} = \sum\limits_{i = 1}^{n \times n} {{\mathop {{y_{ji}}}\limits^ {\sim}}{r_{ji}}}\\
	%{r_{ji}} = \frac{1}{{2\pi}}{e^{ - {{d_{ji}^2}}}},
	%\end{array} \right.
	\left\{ \begin{array}{l}
		\begin{aligned}
			{V_{{N_{ij}}}} &= \sum\limits_{q = 1}^{n \times n} {\mathop {{y_q}}\limits^{\sim} {r_q}} \\
			{r_q} &= \frac{1}{{2\pi {{\sigma}_q} }}{e^{ - \frac{{d_q^2}}{{{\sigma_q^2}}}}},
		\end{aligned}
	\end{array} \right.
	\left\{ \begin{array}{l}
		\begin{aligned}
			{\mathop {{y_{q}}}\limits^ {\sim}} &= 1, &if \mathop {{y_q}}\limits^ \wedge = 1\\
			{\mathop {{y_{q}}}\limits^ {\sim}} &= -1, &if \mathop {{y_q}}\limits^ \wedge = 0,
		\end{aligned}
	\end{array} \right.
\end{equation}
where $N_{ij}$ represents the $i$th row and $j$th column pixel and ${V_{{N_{ij}}}}$ is the total reliability of pixel $N_{ij}$. $d_{q}$ is the distance measure between pixel $N_{ij}$ and center pixel of the $q$th patch $\mathbb{N}_q$. ${r_q}$ is the reliability of the pixel $N_{ij}$ in the patch $\mathbb{N}_q$. ${\mathop {{y_{q}}}\limits^ \wedge}$ represents the prediction of the $q$th patch $\mathbb{N}_q$ and $\mathop {{y_{q}}}\limits^ {\sim}$ represents the class of the prediction. $\sigma_q$ is variance in the patch $\mathbb{N}_q$. Finally, the final binary change map is obtained by determining the threshold.

% \usepackage{multirow}

\begin{figure*}[htb]
	\begin{minipage}[b]{1.0\linewidth}
		\includegraphics[width=\textwidth]{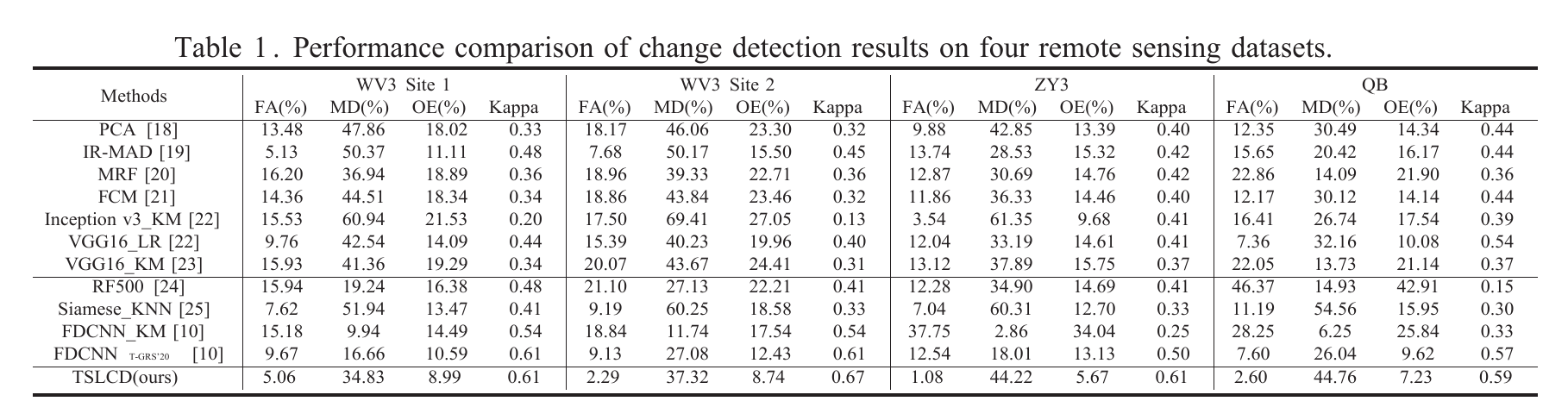}
	\end{minipage}
	\vspace{-0.8cm}
	\label{fig:res12}
\end{figure*}

\vspace{-0.2cm}
\section{EXPERIMENTAL ANALYSIS}
\label{sec:experimental}
\vspace{-0.2cm}
\subsection{Datasets and Implementation Details}
\label{ssec:datset}

%In order to evaluate the effectiveness of the proposed TSLCD method, we conduct the experiments on four publicly available datasets \cite{9052762} from different sensors. WV3 Site 1 and WV3 Site 2 datasets are captured by the WorldView-3 satellite in 2010 and 2015, respectively. And They are captured from different areas in Shenzhen, China, with a size of 1431 × 1431 pixels and a spatial resolution of 2 m. The Zi-Yuan 3 (ZY3) and the QuickBird (QB) datasets are collected by Zi-Yuan 3 satellite and QuickBird satellite, respectively, and they are obtained from different areas in Wuhan, China. The size and spatial resolution of ZY3 are 458 × 559 pixels and 5.8 m, respectively. The first image was captured in 2009 and the second image was captured in 2014. The size and spatial resolution of QB are 1154 × 740 pixels and 2.4 m, respectively. The first image was captured in 2014 and the second image was captured in 2016. In the self-supervised learning phase, the dataset is divided into 2,000,000 patches and 50,000 patches.

In order to evaluate the effectiveness of the proposed TSLCD method, we conduct the experiments on four publicly available datasets \cite{9052762} from different sensors. The size of WV3 Site 1 and WV3 Site 2 datasets is 1431 × 1431. The size of ZY3 dataset is 458 × 559 and the size of QB dataset is 1154 × 740. In the train phase, we use pseudo-labeled samples generated by the traditional CVA \cite{saha2019unsupervised} to train the model. The $n$ is set as 7. The $k$ is set as 32 and $\gamma$ is set as 15. For the change detection methods, we compare the proposed TSLCD with unsupervised methods (PCA \cite{celik2009unsupervised}, IR-MAD \cite{nielsen2007regularized}, MRF \cite{bruzzone2000automatic}, FCM \cite{ghosh2011fuzzy}, Inception v3 \cite{pomente2018sentinel}, VGG16\_LR \cite{pomente2018sentinel}, VGG16\_KM \cite{hou2017change}) and supervised methods (RF500 \cite{ham2005investigation}, Siamese\_KNN \cite{zhan2017change}, FDCNN\_KM \cite{9052762}, FDCNN \cite{9052762}). To evaluate the quantitative accuracy of the estimated result, four objective evaluation metrics including Overall Error (OE) , Kappa coefficient (Kappa), Missed Detection (MD) and False Alarm (FA) are employed.

%(a) $\bm{X}_1$ obtained in 2010.
%(b) $\bm{X}_2$ obtained in 2015. (c) GT of (a) and (b). Change maps produced by (d) PCA, (e) IR-MAD, (f) MRF, (g) VGG16\_KM, (h) FDCNN, (i) TSLCD.
\vspace{-0.2cm}
\subsection{Experiment Results}
\label{ssec:result}

In this section, we perform experiments on four available datasets and compare the performance of our proposed method with the state-of-the-arts. In Table 1, compared with other approaches, the proposed TSLCD achieves the best results with the lowest OE and highest Kappa. Because the task-related self-supervised learning module can extract task-consistent features and smooth mechanism can remove some pseudo-changes by exploiting spatial neighborhood information.

In order to prove the effectiveness of the proposed TSLCD method, the experimental results of all methods in WV Site 2 dataset are illustrated in Fig. 2. By observing the details of the results, it can be seen that TSLCD without Smooth Mechanism (SM) has a better ability to detect changes and SM has a great suppression effect on pseudo-changes and noise.

\begin{figure}[htb]
	\begin{minipage}[b]{1.0\linewidth}
		\includegraphics[width=\textwidth]{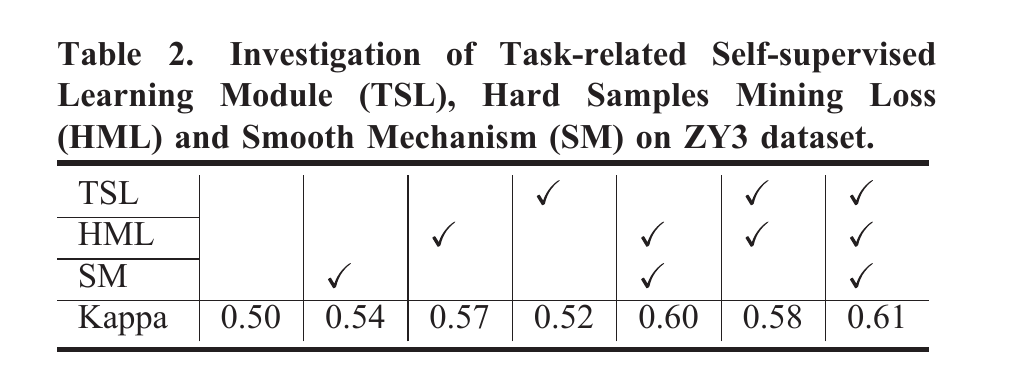}
	\end{minipage}
	\vspace{-0.8cm}
	\label{fig:res12}
\end{figure}

In addition, the algorithms such as FDCNN, VGG-16\_KM, FCM and MRF can bring a lot of false detection. This result demonstrates that the self-supervised learning module can extract the features of original image effectively. Due to the limited ability of the network to mine hard samples, the designed loss function increases the weight of the hard samples to increase its contribution to the loss, where the network can exploit the hard samples better. Finally, the smooth mechanism is utilized to smooth the change map to reduce the noise. At last, Table 2 proves the effectiveness of different modules in our proposed method. In conclusion, the different experiments on four datasets can clearly verify that the proposed TSLCD has better performance of change detection and pseudo-changes suppression.
\vspace{-0.2cm}
\begin{figure}[htb]
	
	\begin{minipage}[b]{.32\linewidth}   %.32
		\centering
		\centerline{\includegraphics[width=2.4cm]{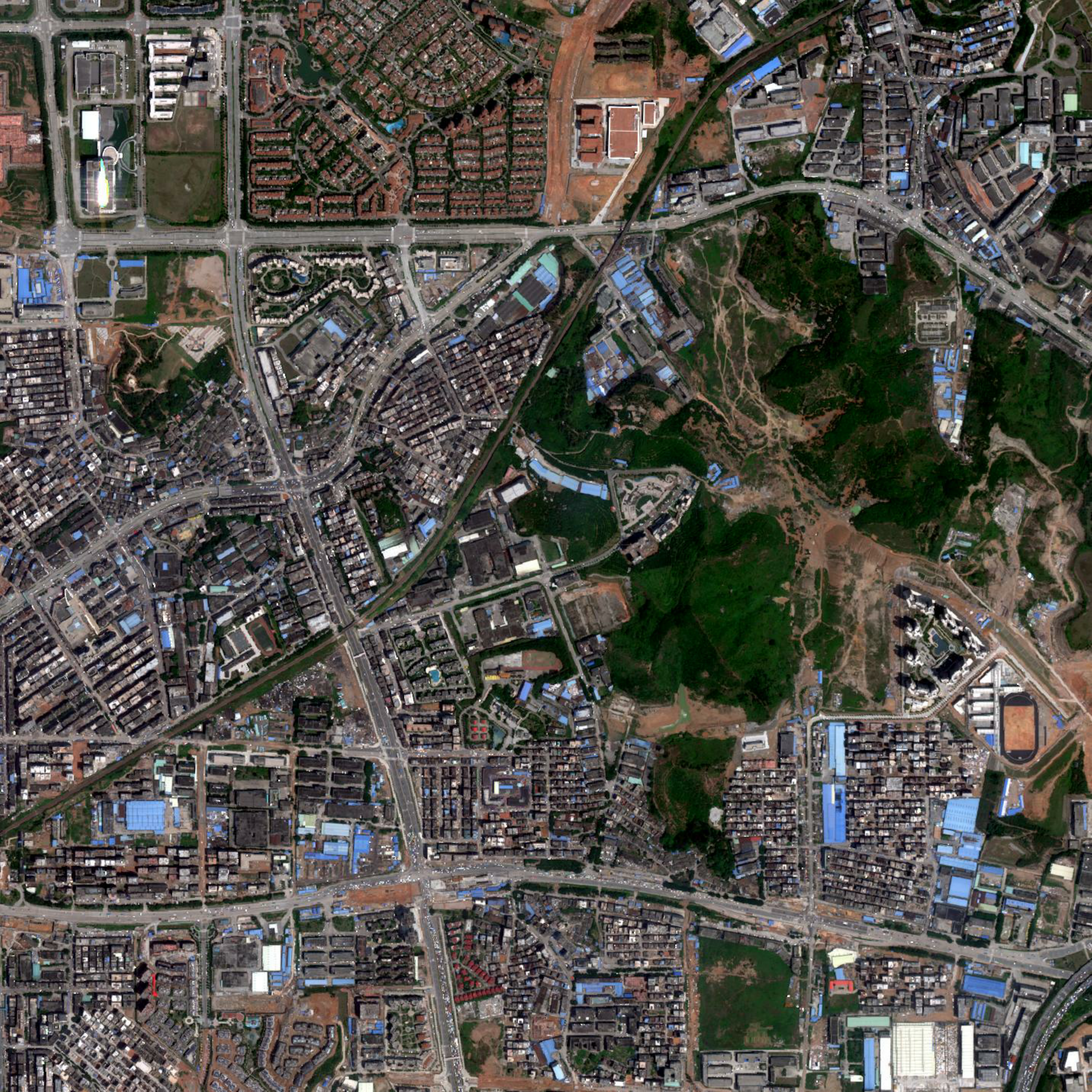}}
		\vspace{-2.0mm}
		\centerline{\scriptsize (a) Prechange Image}\medskip
		\vspace{-2.0mm}
	\end{minipage}
	%\hfill
	\begin{minipage}[b]{.32\linewidth}
		\centering
		\centerline{\includegraphics[width=2.4cm]{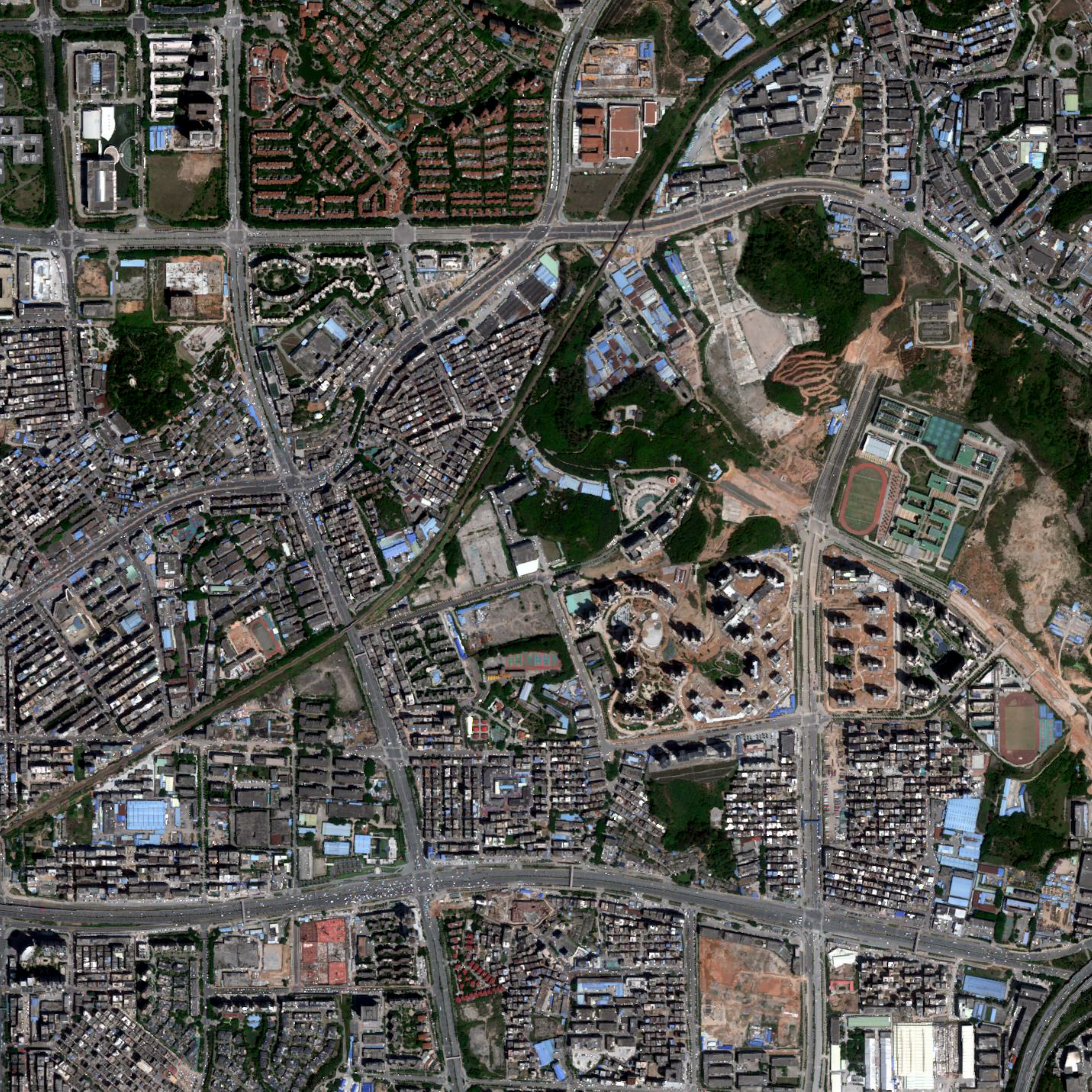}}
		\vspace{-2.0mm}
		\centerline{\scriptsize (b) Postchange Image}\medskip
		\vspace{-2.0mm}
	\end{minipage}
	%\hfill
	\begin{minipage}[b]{.32\linewidth}
		\centering
		\centerline{\includegraphics[width=2.4cm]{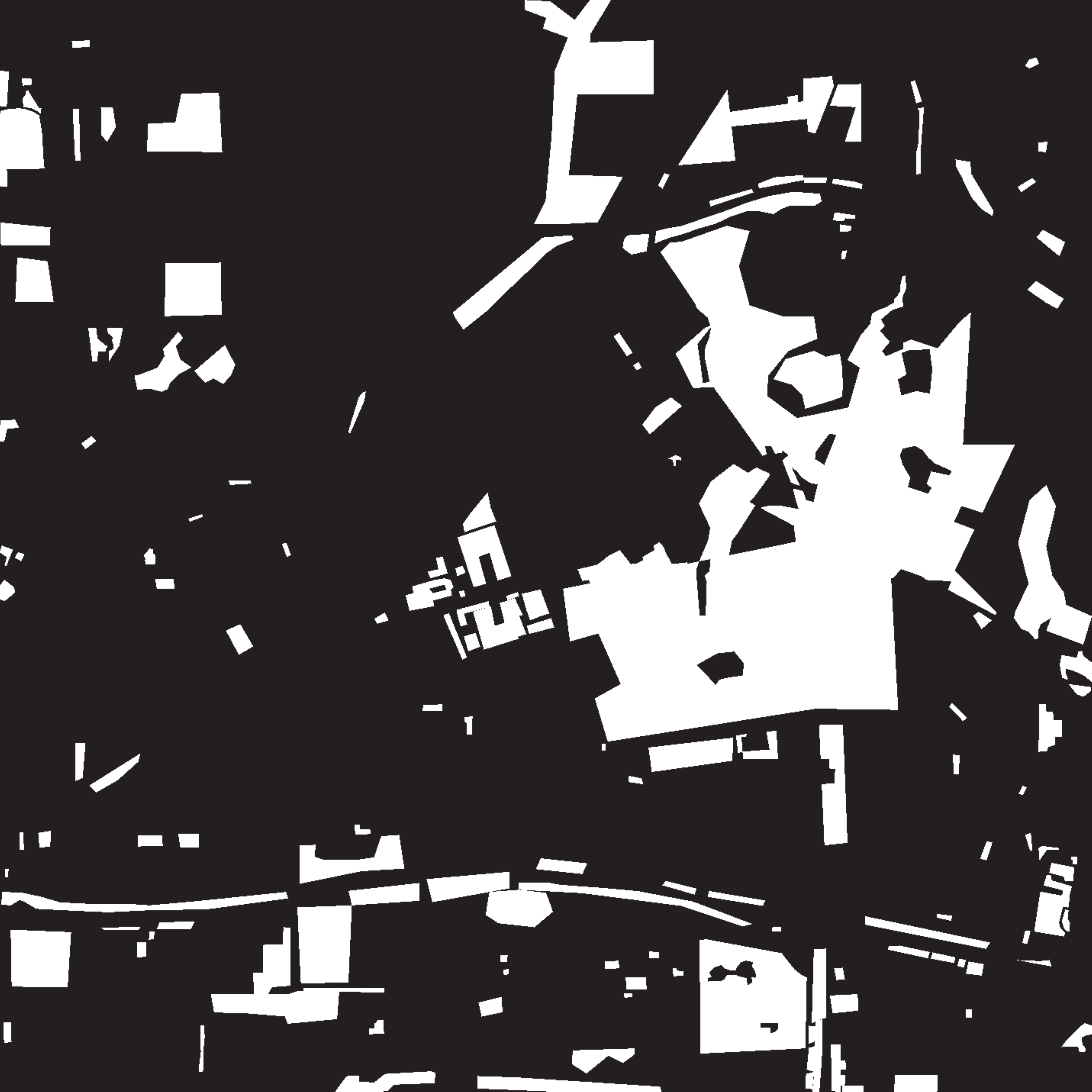}}
		\vspace{-2.0mm}
		\centerline{\scriptsize (c)  GT}\medskip
		\vspace{-2.0mm}
	\end{minipage}
	
	%\hfill	
	%	\begin{minipage}[b]{.32\linewidth}
	%		\centering
	%		\centerline{\includegraphics[width=2.5cm]{pca}}
	%		\vspace{-2.0mm}
	%		\centerline{\scriptsize (d) PCA \cite{celik2009unsupervised}}\medskip
	%		\vspace{-2.0mm}
	%	\end{minipage}
	%	\hfill %
	\begin{minipage}[b]{.32\linewidth}
		\centering
		\centerline{\includegraphics[width=2.4cm]{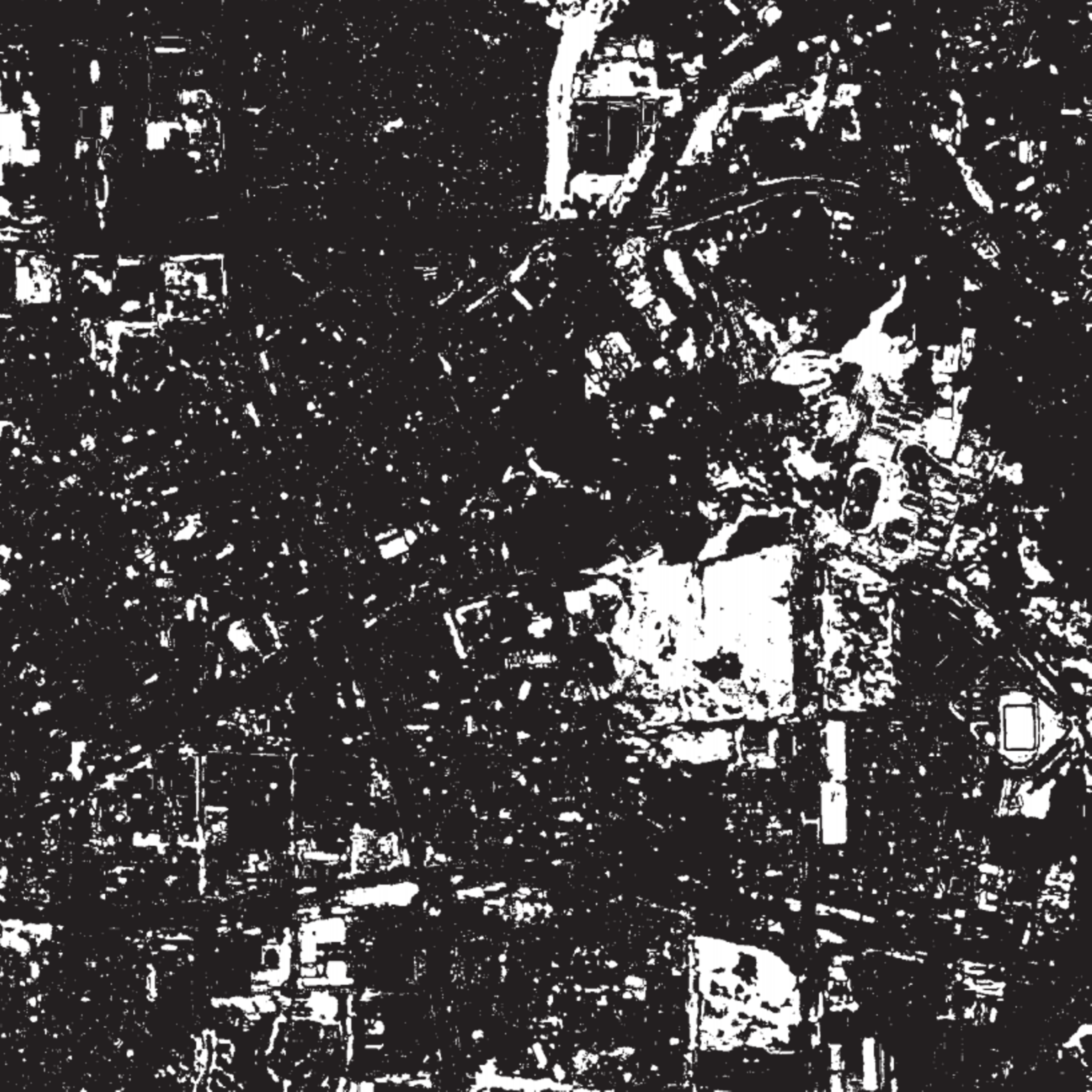}}
		\vspace{-2.0mm}
		\centerline{\scriptsize (d) IR-MAD \cite{ghosh2011fuzzy}}\medskip
		\vspace{-2.0mm}
	\end{minipage}
	%\hfill
	\begin{minipage}[b]{0.32\linewidth}
		\centering
		\centerline{\includegraphics[width=2.4cm]{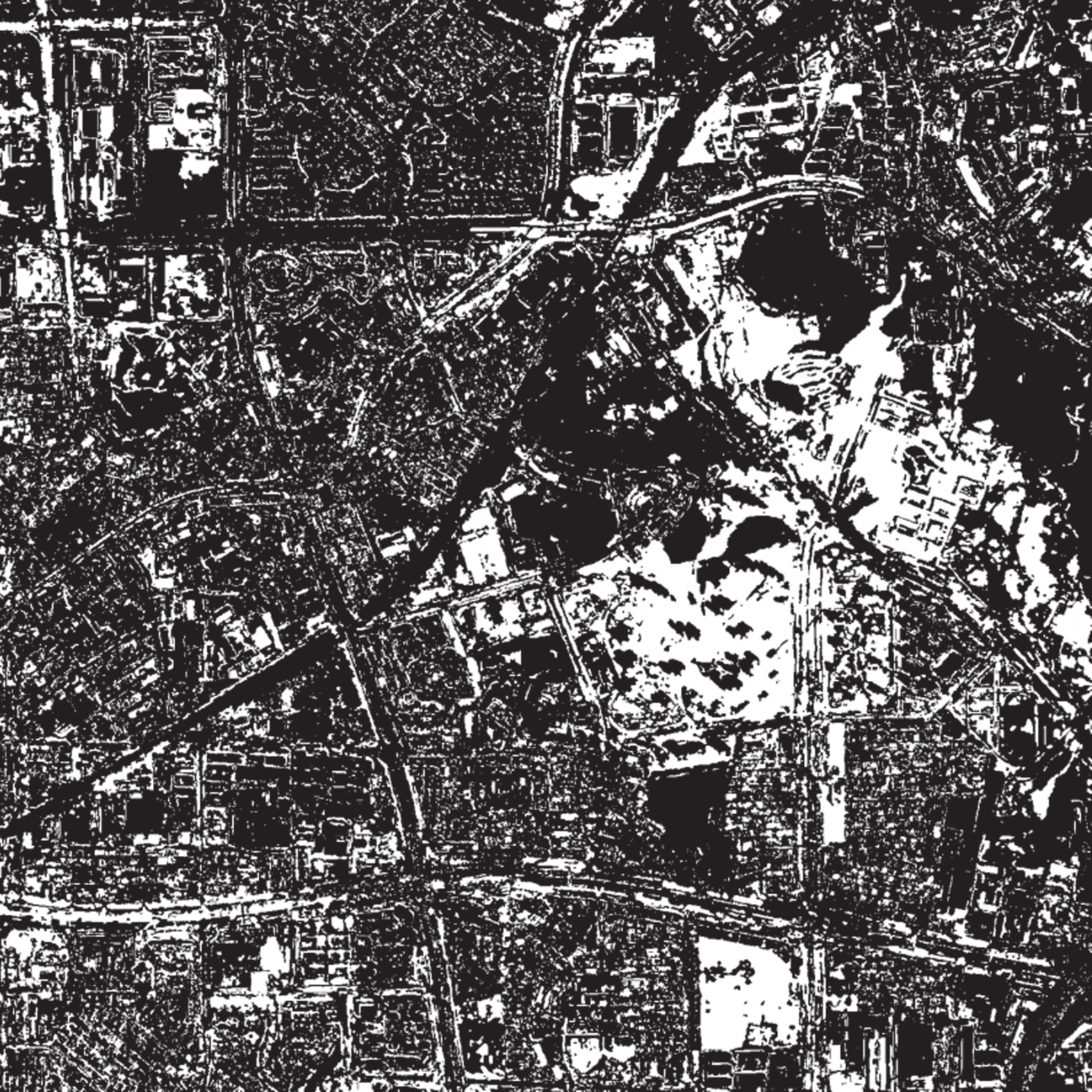}}
		\vspace{-2.0mm}
		\centerline{\scriptsize (e) MRF \cite{bruzzone2000automatic}}\medskip
		\vspace{-2.0mm}
	\end{minipage}
	%\hfill
	\begin{minipage}[b]{.32\linewidth}
		\centering
		\centerline{\includegraphics[width=2.4cm]{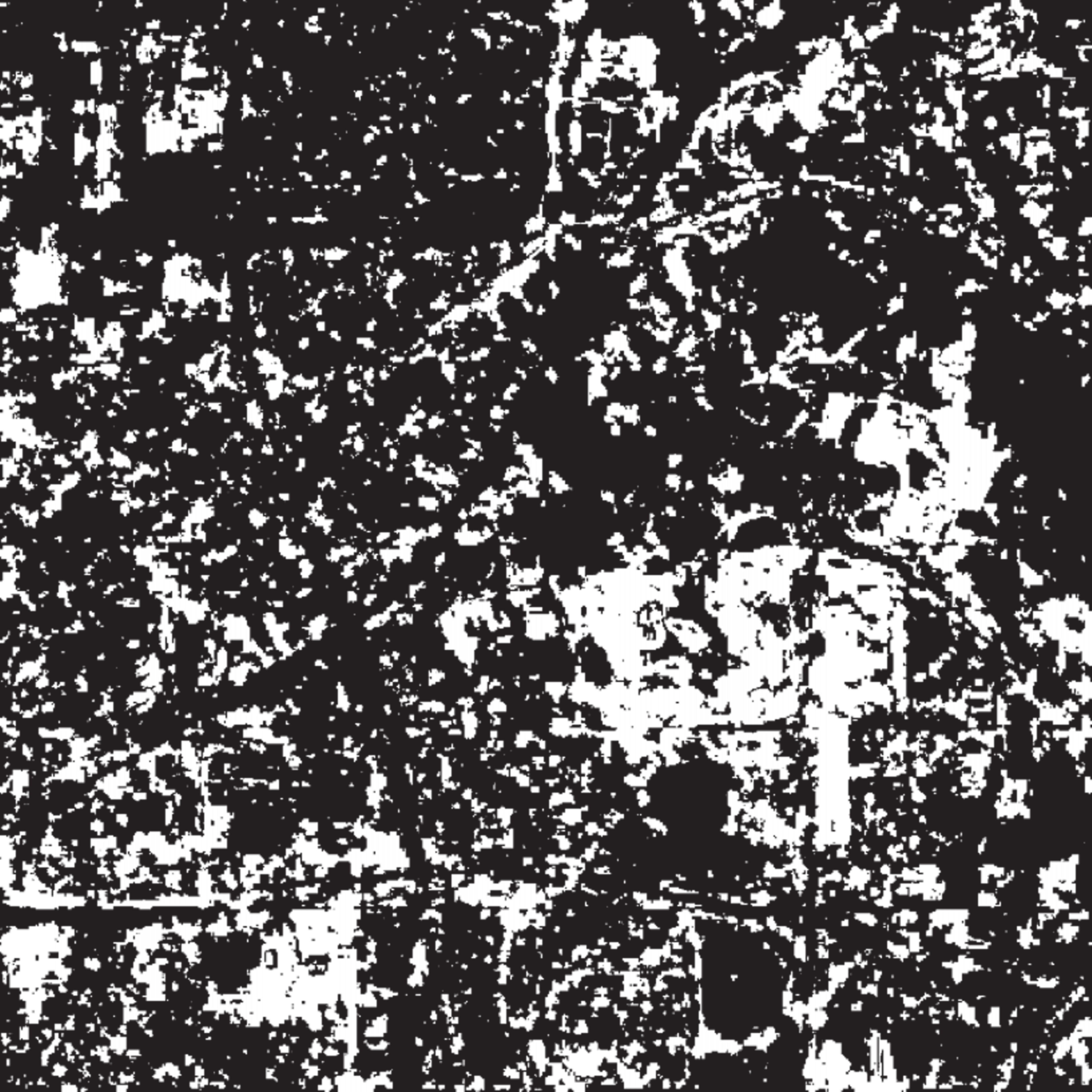}}
		\vspace{-2.0mm}
		\centerline{\scriptsize (f) VGG16\_KM \cite{hou2017change}}\medskip
		\vspace{-2.0mm}
	\end{minipage}
	%\hfill
	
	\begin{minipage}[b]{.32\linewidth}
		\centering
		\centerline{\includegraphics[width=2.4cm]{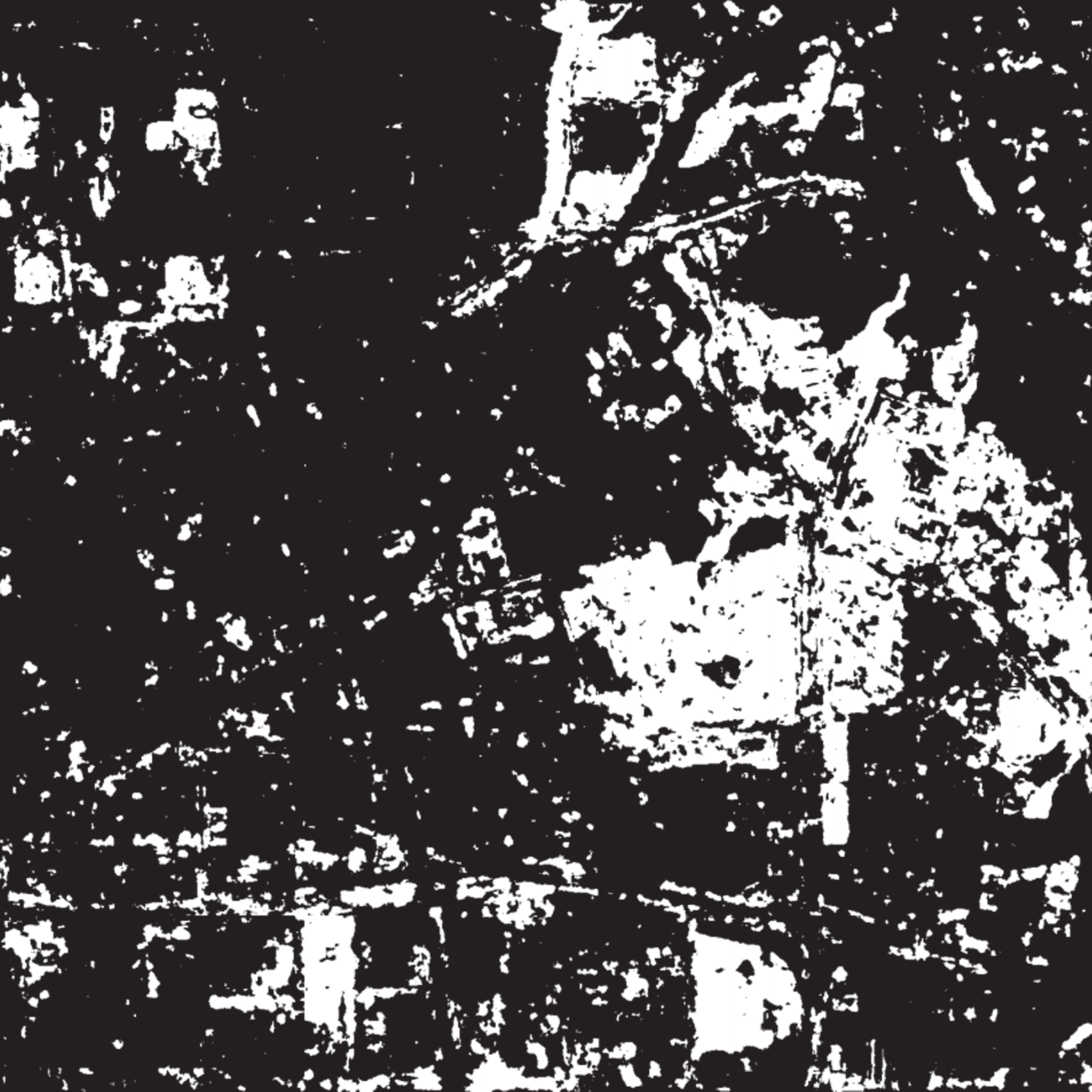}}
		\vspace{-2.0mm}
		\centerline{\scriptsize (g) FDCNN \cite{9052762}}\medskip
		\vspace{-2.0mm}
	\end{minipage}
	%\hfill
	\begin{minipage}[b]{.32\linewidth}
		\centering
		\centerline{\includegraphics[width=2.4cm]{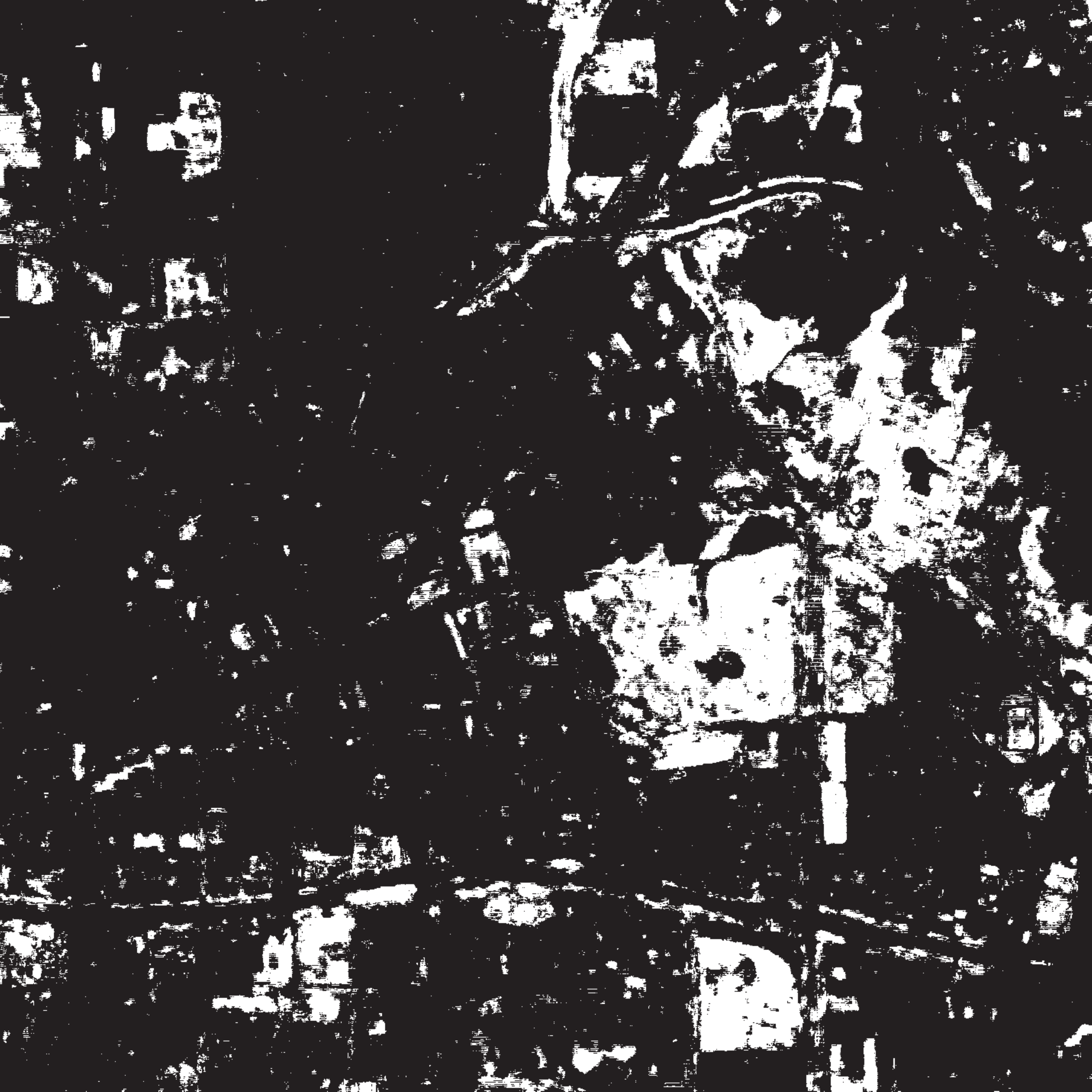}}
		\vspace{-2.0mm}
		\centerline{\scriptsize (h) TSLCD (w/o SM)}\medskip
		\vspace{-2.0mm}
	\end{minipage}
	%\hfill
	\begin{minipage}[b]{.32\linewidth}
		\centering
		\centerline{\includegraphics[width=2.4cm]{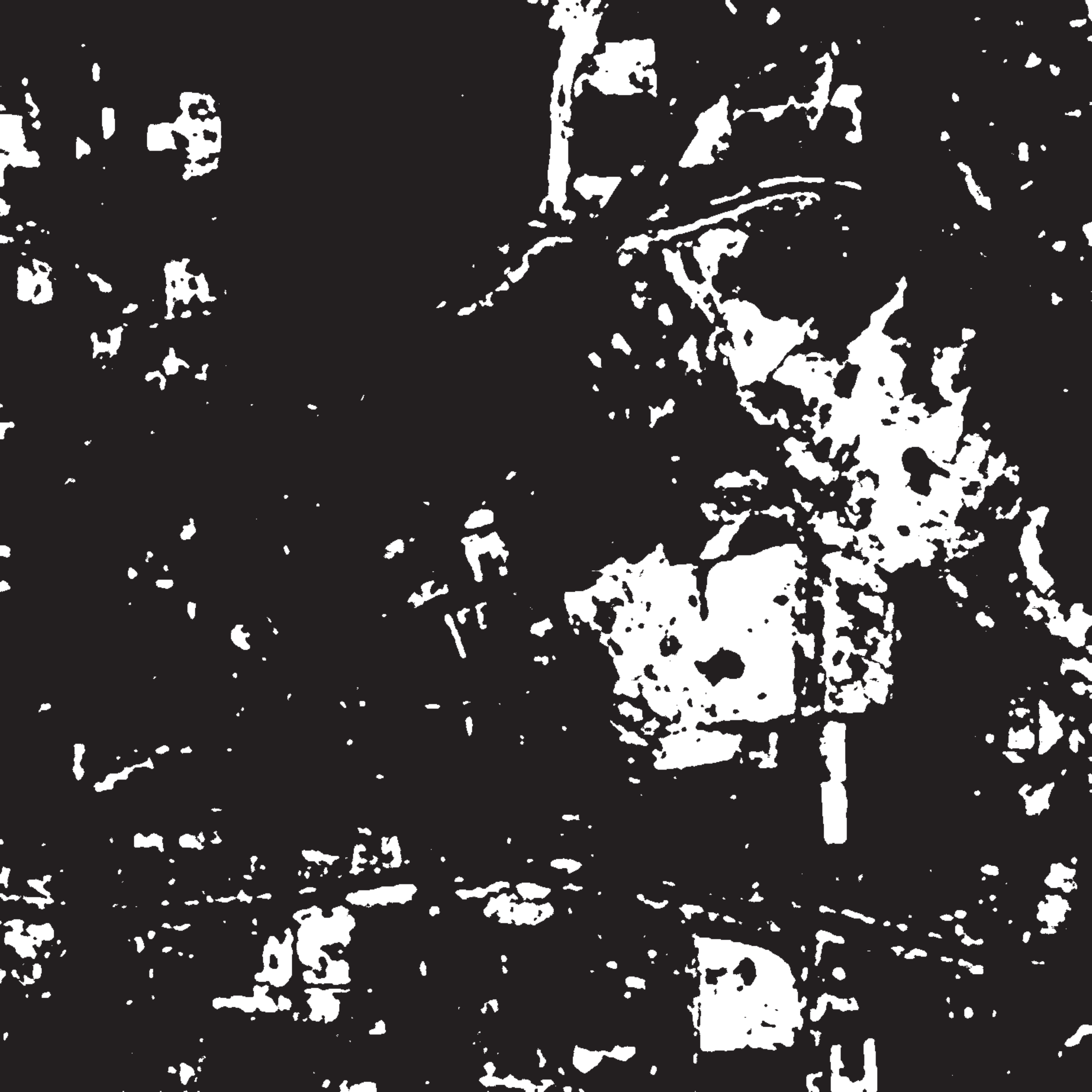}}
		\vspace{-2.0mm}
		\centerline{\scriptsize (i) TSLCD}\medskip
		\vspace{-2.0mm}
	\end{minipage}
	%\hfill 
	\vspace{-0.1cm}
	\caption{Change detection results on the WV3 Site 2 testing dataset. (a) Prechange Image. (b)Postchange Image. (c) GT. (d-g) Competitors. (h) TSLCD (without SM). (i) TSLCD.
	}
	\label{fig:res3}
	\vspace{-0.75cm}
\end{figure}

\section{CONCLUSION}
\label{sec:con}

This work introduces an unsupervised change detection approach based on TSLCD, in which the task-related self-supervised learning network is used to learn better features from remote sensing images, a hard-sample-mining loss function is utilized to pay more attention to hard-to-classify samples. Moreover, a smooth mechanism has been proposed for reducing pseudo-changes and noise and make it more effective to obtain the final binary change map. Experiments on four datasets from different sensors verify that the proposed TSLCD approach achieves state-of-the-art performance.

\bibliographystyle{IEEEbib}
\bibliography{icme2021template}
\end{document}